\documentclass[twoside]{article}
\usepackage[a4paper]{geometry}
\usepackage[latin1]{inputenc} 
\usepackage[T1]{fontenc} 
\usepackage{RR}
\usepackage{hyperref}
\usepackage{graphics} 
\usepackage{graphicx}
\usepackage{epsfig} 
\usepackage{subfigure}
\usepackage{amsmath} 
\usepackage{amssymb}  
\usepackage{latexsym}

\RRNo{8702}
\RRdate{March 2015}
\RRauthor{
Konstantin E. Avrachenkov\thanks{K.E. Avrachenkov is with Inria Sophia Antipolis, 2004 Route des Lucioles, 06902,
Sophia Antipolis, France {\tt\small k.avrachenkov@inria.fr}}%
  \and
Vivek S. Borkar\thanks{V.S. Borkar is with the Department of Electrical Engineering, IIT Bombay,
Powai, Mumbai 400076, India {\tt\small borkar.vs@gmail.com}}%
  \thanks[sfn]{This work was partially supported by Indo-French CEFIPRA Collaboration Grant
No.5100-IT1 ``Monte Carlo and Learning Schemes for Network Analytics''.}%
}
\authorhead{K.E. Avrachenkov \& V.S. Borkar}
\RRtitle{Index de Whittle pour Crawling du Contenu \'Eph\'em\`ere}
\RRetitle{Whittle Index Policy for\\ 
Crawling Ephemeral Content}
\titlehead{Whittle Index Policy for Crawling Ephemeral Content}
\RRresume{
Nous consid\'erons une t\^ache de la planification du parcours d'un robot pour 
r\'ecup\'erer le contenu \'eph\'em\`ere de plusieurs sites web. 
Typiquement, un utilisateur de web perd int\'er\^et pour le contenu \'eph\'em\`ere,
comme des nouvelles ou des posts aux r\'eseaux sociaux, apr\`es plusieurs jours ou 
m\^eme heures.
Donc, le d\'eveloppement de la planification dynamique du parcours de 
ces sources d'information \'eph\'em\`eres est tr\`es important. 
Nous formulons d'abord ce probl\`eme comme un probl\`eme de commande optimale
avec une r\'ecompense moyenne. La r\'ecompense peut \^etre mesur\'ee par
le nombre de clics ou par le nombre de demandes de recherche pertinente. 
Le probl\`eme dans sa formulation initiale souffre de la ``mal\'ediction
de la dimension'' et devient rapidement inextricable m\^eme avec nombre mod\'er\'e
des sources d'information. Heureusement, ce probl\`eme admet un Index de Whittle, 
qui conduit \`a la d\'ecomposition du probl\`eme et \`a une politique de parcours 
tr\`es simple et efficace. 
Nous d\'erivons l'Index de Whittle et fournissons sa justification th\'eorique.
}
\RRabstract{
We consider a task of scheduling a crawler to retrieve content from several sites
with ephemeral content. A user typically loses interest in ephemeral content,
like news or posts at social network groups, after several days or hours.
Thus, development of timely crawling policy for such ephemeral information sources
is very important. We first formulate this problem as an optimal control problem
with average reward. The reward can be measured in the number of clicks or relevant
search requests. The problem in its initial formulation suffers from the curse
of dimensionality and quickly becomes intractable even with moderate number
of information sources. Fortunately, this problem admits a Whittle index, which
leads to problem decomposition and to a very simple and efficient crawling policy.
We derive the Whittle index and provide its theoretical justification.
}
\RRmotcle{Index de Whittle, Moteur de Recherche, Crawler, Robot, Contenu \'Eph\'em\`ere}
\RRkeyword{Whittle Index, Search Engine, Crawler, Ephemeral Content} 

\RRprojet{Maestro}  
\RCSophia 

\begin{document}
\makeRR   

\section{Introduction}

Nowadays an overwhelming majority of people find new information on the web at news sites,
blogs, forums and social networking groups. Moreover, most information consumed is
ephemeral in nature, that is, people tend to lose their interest in the content
in several days or hours. The interest in a content can be measured in terms of clicks
or number of relevant search requests. It has been demonstrated that the interest
decreases exponentially over time \cite{GBL10,LOSS11,Moonetal11}.

In a series of works (see e.g., \cite{CGM00,CN02,CGM03,ON10} and references therein)
the authors address the problem of refreshing documents in a database. However, these works
do not consider the ephemeral nature of the information. Motivated by this challenge,
the authors of \cite{LOSS11} suggest a procedure for optimal crawling of ephemeral
content. Specifically, the authors of \cite{LOSS11} formulate an optimization problem
for finding optimal frequencies of crawling for various information sources.

The approach presented in \cite{LOSS11} is static, in the sense that
the distribution of crawling effort among the content sources is always the same independent
of the time epoch and, in particular, does not depend on any `state variable(s)'
evolving with time. With a dynamic policy, for instance, if there is not much new material
on the principal information sources, the crawler could spend some time to crawl the
sources with less popular content but which nevertheless bring noticeable rate of clicks
or increase information diversity. Therefore, in the present work we suggest a dynamic
formulation of the problem as an optimal control problem with average reward.
The direct application of dynamic programming quickly becomes intractable even
with moderate number of information sources, due to the so-called curse of dimensionality.
Fortunately, the problem admits a Whittle index, which leads to problem decomposition and
to a very simple and efficient crawling policy. We derive the Whittle index and provide
its theoretical justification.

In \cite{ADKNS11,LN06,TLNC01} the authors study the interaction between the crawler and the indexing engine
by means of optimization and control theoretic approaches. One of interesting future research
directions is to take into account the indexing engine dynamics in the present context.

The general concept of the Whittle index was introduced by P. Whittle in \cite{Whittle}.
This has been a very successful heuristic for restless bandits, which, while suboptimal in
general, is provably optimal in an asymptotic sense \cite{Verloop15,Weber} and has good empirical performance.
It and its variants have been used extensively in logistical and engineering applications,
some recent instances of the latter in communications and control being for sensor scheduling \cite{Nino},
multi-UAV coordination \cite{Ny}, congestion control \cite{AADJ13,AHPZ14,JS07},
channel allocation in wireless networks \cite{LAV15}, cognitive radio \cite{Liu}
and real-time wireless multicast \cite{Raghu}.
Book length treatments of indexable restless bandits appear in \cite{J10,Ruiz}.


\section{Model}

There are $N$ sources of ephemeral content. A content at source $i \in \{1,...,N\}$
is published with an initial utility modelled by a nonnegative random variable $\xi_i$ and decreasing
exponentially over time with a deterministic rate $\mu_i$. The new content arrives
at source $i \in \{1,...,N\}$ according to a time-homogeneous Poisson process with rate $\Lambda_i$.
Thus, if source $i$'s content is crawled
$\tau$ time units after its creation, its utility is given by $\xi_i \exp(-\mu_i\tau)$.
The base utility $\xi_i$ is assumed independent identically distributed across contents at a given
source, with a finite mean $\bar{\xi}_i$. It is also assumed independent across sources.
We assume that  the crawler crawls periodically at multiples of time $T > 0$ and has to choose at each
such instant which sources to crawl, subject to a constraint we shall soon specify.
When the crawler
crawls a content source, we assume that the crawling is done in an exhaustive manner.
In such a case, the crawler obtains the following expected reward from crawling
the content of source $i$:
\begin{equation}
\label{eq:expreward}
u_i = \Lambda_i E[\xi_i\exp(-\mu_i \tau)] = \frac{\Lambda_i \bar{\xi}_i}{\mu_i}(1-\exp(-\mu_i T)).
\end{equation}
Set $\alpha_i = \exp(-\mu_i T)$. Let us define the state
of source $i$ at time $t$  as the total expected utility of its content, denoted by $X_i(t)$.
Then, if we do not crawl source $i$ at epoch $t$ (formally, the control is $v_i(t)=0$ - we
say the source is `passive'), we obtain zero reward $r_i(X_i(t), v_i(t)) = 0$ and the state evolves as follows:
\begin{equation}
\label{eq:Xv0}
X_i(t+1) = \alpha_i X_i(t) + u_i.
\end{equation}
On the other hand, if we crawl source $i$ (formally, $v_i(t)=1$ - we say the source is `active'), we obtain
the expected reward $r_i(X_i(t), v_i(t)) = X_i(t)$ and the next state of the source is given by
\begin{equation}
\label{eq:Xv0a}
X_i(t+1) = u_i.
\end{equation}

Our aim is to maximize the long run average reward
\begin{equation}
\limsup_{t \uparrow \infty}\sum_{i=1}^N\frac{1}{t}\sum_{m = 0}^t r(X_i(t), v_i(t)) \label{reward}
\end{equation}
subject to the constraint
\begin{equation}
\limsup_{t \uparrow \infty}\sum_{i=1}^N\frac{1}{t}\sum_{m = 0}^t C_iv_i(t) = M \label{constraint}
\end{equation}
for a prescribed $M > 0$. If $C_i=1, i=1,...,N$,
this case can be interpreted as a constraint on the number of crawled sites per
crawling period $T$ and corresponds to the original Whittle framework for restless bandits \cite{Whittle}.
The case $C_i \neq 1$ is slightly more general and can represent the situation when various sites have
different limits on the crawling rates (typically specified in the file `robots.txt').


This is a constrained average reward control problem \cite{Altman,P97}. We address this problem in the
framework of restless bandits and derive a simple index policy for the problem, which may be viewed
as a variant of the celebrated Whittle index. In the next section, we recall the theory of Whittle index.\\


\section{Whittle index}

The original formulation of restless bandits is for discrete state space Markov chains, but we consider here
Markov chains with  closed domains (i.e., closure of an open set)  $S_i \subset \mathcal{R}^d, d \geq 1$, as
state space.
The original motivation for the index policy remains valid nevertheless as
long as we justify the associated dynamic programming equation, which we do. A deterministic dynamics such as ours is
a special case, albeit degenerate. The fully stochastic case can be handled similarly
and is detailed in the report \cite{RR8702}. While we introduce the broader framework
in a general set up, we use the same notation as above to highlight the correspondences. This should not cause
any confusion.\\

Thus consider resp.\ $S_i$-valued processes $X_i(t), t \geq 0, 1 \leq i \leq N$, each with
two possible dynamics, dubbed
active and passive, wherein they are governed by transition kernels $p_i(dy|x), q_i(dy|x)$ resp. These are
assumed to be continous as maps $x \in S_i \mapsto \mathcal{P}(S_i)$. ($:=$ the space of probability measures
on $S_i$ with Prohorov topology). The control at time $t$ is an $A := \{0, 1\}^N$-valued vector
$v(t) = [v_1(t), \cdots, v_N(t)] \in A$, with the understanding
that $v_i(t) = 1 \Longleftrightarrow X_i(t)$ is active. In the original restless bandit problem, exactly
$N' < N$ processes are active at
any given time. The $v_i(t)$ are assumed to be adapted to the history, i.e., the $\sigma$-field $\sigma(X_i(s), s \leq t; v_i(s), s < t; 1 \leq i \leq N)$. Let $r_i : S \mapsto \mathcal{R}^+, 1 \leq i \leq N,$
be reward functions so that a reward of $r_i(X_i(t))$ is
accrued if process $i$ is active at time $t$. The objective then is to maximize the long run average reward
\begin{displaymath}
\limsup_{t \uparrow \infty}\sum_{i = 1}^N\frac{1}{t}\sum_{m = 0}^tE[r_i(X_i(t))v_i(t)].
\end{displaymath}

This problem has state space $\times_{i=1}^N S_i$. Whittle's heuristic among other things reduces the problem to
separate control problems on $S_i$. The idea is to relax the constraint of `exactly $N'$ are active' to `on the
average, $N'$ are active', i.e., to
\begin{displaymath}
\limsup_{t \uparrow \infty} \frac{1}{t}\sum_{s = 0}^tE[\sum_{i = 1}^Nv_i(s)] = N'.
\end{displaymath}
This makes it a constrained average reward control problem \cite{Altman,P97} which permits a relaxation
to an unconstrained
average reward problem by replacing the above reward by
\begin{displaymath}
\limsup_{t \uparrow \infty}\sum_{i = 1}^N\frac{1}{t}\sum_{s = 0}^tE[r_i(X_i(s))v_i(s)
+ \lambda (N'/N - v_i(s))],
\end{displaymath}
where $\lambda \in \mathcal{R}$ is the Lagrange multiplier. Motivated by this, Whittle
introduced a `subsidy' $\lambda$ for passivity, i.e., a
virtual reward for a process in passive mode. Replace the above control problem by $N$ control problems with
the $i$th problem for process $X_i(\cdot)$ seeking to maximize over admissible $v_i(t), t \geq 0$, the reward
\begin{equation}
\limsup_{t \uparrow \infty}\frac{1}{t}\sum_{s = 0}^tE[r_i(X_i(t))v_i(s) + \lambda(N'/N - v_i(s))].
\label{reward2}
\end{equation}
The dynamic programming equation for this average reward problem is
$$
V_i(x) + \beta  =
$$
\begin{equation}
\max\Big(\lambda + \int q_i(dy|x)V_i(y), \ r_i(x) + \int p_i(dy|x)V_i(y)\Big). \label{DP}
\end{equation}
If this can be rigorously justified (which is not always easy), one defines
$B(\lambda)$ as the set of passive states, i.e.,
$$
B(\lambda) :=
$$
\begin{displaymath}
\left\{ x : \lambda + \int q_i(dy|x)V_i(y) \geq r_i(x) + \int p_i(dy|x)V_i(y)\right\}.
\end{displaymath}
If $B(\lambda)$ increases monotonically from $\phi$ to $S_i$
as $\lambda$ increases from $-\infty$ to $\infty$, the problem is said to be
\textit{Whittle indexable}. The Whittle index for the $i$th process in state $x_i$ is then defined as

\newpage

$$
\gamma_i(x_i) :=
$$
\begin{displaymath}
\{ \lambda' : \lambda' + \int q_i(dy|x_i)V(y) = r_i(x_i) + \int p_i(dy|x_i)V(y)\}.
\end{displaymath}
The so-called `\textit{Whittle index policy}' \cite{Whittle} then is to set $v_i(t) = 1$ for the $i$ with the top $N'$ indices and $v_j(t) = 0$ for the rest.\\

\section{Dynamic programming equation}

In view of the above, the first step is to justify the counterpart of (\ref{DP}) in our context. For this, we first note that
$r_i(x) = x, 1 \leq i \leq N$. Further, let $u_i^* := \frac{u_i}{1 - \alpha_i} > u_i$. We argue that without
loss of generality, we may take $S_i = [u_i, u^*_i]$. To see this, let $X_i(0) = x_0$. If $x_0 \leq u^*_i$,
it is easy to see that
\begin{displaymath}
X_i(t) \ \leq \ \alpha_i^tx_0 + (1 - \alpha_i^t)u_i^* \ \uparrow \ u^*_i,
\end{displaymath}
where the equality in the first inequality occurs only if source $i$ is never crawled. On the other hand,
if $x_0 > u^*_i$, then
\begin{displaymath}
X_i(t) = \alpha^t_ix_0 + (1 - \alpha_i^t)u^*_i \downarrow  u^*_i \ \mbox{as} \ t \uparrow \infty,
\end{displaymath}
if never crawled, and reduces to the previous case if there is even a single crawl.
Combining the two observations and recalling
that we consider the long-run average criterion, we conclude that
$x_0 \notin [u_i, u_i^*]$ are transient and can be ignored. Thus we set $S_i = [u_i, u^*_i]$.\\

Henceforth we focus on the average reward problem for source $i$. We do not delve into the justification
for Lagrange multiplier formulation for constrained average cost problem on a general state space, as this is well understood. (In fact, it follows from standard Lagrange multiplier theory applied to the `occupation measure' formulation of average cost problem which casts it as an abstract linear program. See section 4.2 of \cite{Borkar} which carries out this program for discrete state space and section 3.2 of \textit{ibid.} which describes how to extend the same to general compact Polish state spaces as long as the controlled transition probability kernel is continuous in the initial state and control.) For notational simplicity we drop the
index $i$ for the time being. We approach the problem by the standard `vanishing discount' argument.
Thus let $0 < \delta < 1$ be a discount factor and for $k(x, v) := xv + C\lambda(1-v)$, consider the
infinite horizon discounted reward
\begin{displaymath}
\sum_{m = 0}^{\infty}\delta^tk(X(t)).
\end{displaymath}
Denote the associated value function by
\begin{displaymath}
V_{\delta}(x) := \sup_{\{v(t)\}, X(0) = x}\left[\sum_{m=0}^{\infty}\delta^tk(X(t), v(t))\right].
\end{displaymath}
Then $V_{\delta}$ satisfies the discounted reward dynamic programming equation
\begin{equation}
V_{\delta}(x) = \max\left(C\lambda + \delta V_{\delta}(\alpha x + u), \  x + \delta V_{\delta}(u)\right). \label{DP-disc}
\end{equation}

\newpage

\noindent \textbf{Lemma 1} The solution of equation (\ref{DP-disc}) has the following properties: \\

\noindent (1) Equation (\ref{DP-disc}) has a unique bounded continuous solution $V_{\delta}$;

\noindent (2) $V_{\delta}$ is Lipschitz uniformly in $\delta \in (0, 1)$;

\noindent (3) $V_{\delta}$  is  monotone increasing and convex.\\

\noindent \textbf{Proof:}  Claim (1) is standard (See Theorem 4.2.3 and bullet 1 in `Notes on $\S 4.2$', Section 4.2, \cite{Lasserre}).  For (2), consider $x \neq x' > x \in S$. Consider processes $X(t), t \geq 0$, and $X'(t), t \geq 0$, with initial conditions $x, x'$ resp.,
both controlled by control sequence $v(t), t \geq 0,$ that is optimal for the former. Then
\begin{eqnarray*}
V_{\delta}(x') - V_{\delta}(x) &\leq& \sum_{t = 0}^{\infty}\delta^m(k(X'(t), v(t)) - k(X(t), v(t)))  \\
&=& \left(\frac{(1 - \alpha\delta)^{\tau}}{1 - \alpha}\right)(x' - x),
\end{eqnarray*}
where $\tau :=$ the time of first crawl ($= \infty$ if never crawled). Interchanging the roles of $x', x$ we get a symmetric inequality, whence it follows that
\begin{displaymath}
|V_{\delta}(x') - V_{\delta}(x)| \ \leq \ \left(\frac{(1 - \alpha\delta)^{\tau}}{1 - \alpha}\right)|x' - x|.
\end{displaymath}
For the first part of (3), take $x' >  x$ as above and let  $X'(t), X(t), t \geq 0,$ be processes generated by a common admissible control sequence $\{v(t)\}$ with initial conditions $x', x$ resp. Then it is easy to check that $X'(t) \geq X(t)$ for all $t$. Therefore
\begin{equation}
\sum_{t=0}^{\infty}\delta^tk(X'(t), v(t)) \geq \sum_{t=0}^{\infty}\delta^tk(X'(t), v(t)). \label{av-ineq}
\end{equation}
Taking supremum over all admissible controls on both sides,   monotonicity of $V_{\delta}$  follows. For convexity, define the finite horizon discounted value function
\begin{displaymath}
V_n(x) = \sup_{\{v(t)\}, X(0) = x}\sum_{t = 0}^n\delta^tk(X(t), v(t)).
\end{displaymath}
Then it satisfies the dynamic programming equation
\begin{displaymath}
V_n(x) = \max\left(C\lambda + \delta V_{n - 1}(\alpha x + u), \ x + \delta V_{n - 1}(u)\right)
\end{displaymath}
for $n \geq 1$ with $V_0(x) = x$. The convexity of $V_n$ for each $n$ then follows by a simple induction. Since $V_{\delta}(x) = \lim_{n\uparrow\infty}V_n(x)$, $V_{\delta}$ is also convex. \hfill $\Box$

\ \\

Define $\bar{V}_{\delta}(x) = V_{\delta}(x) - V_{\delta}(u), \ x \in S$. Then by the above lemma, $\bar{V}_{\delta}$ is bounded Lipschitz, monotone and convex with $\bar{V}_{\delta}(u) = 0$. Also, $(1 - \delta)V_{\delta}(u)$ is bounded.
Using Arzela-Ascoli and Bolzano-Weirstrass theorems, we may pick a subsequence such that  $(\bar{V}_{\delta}, (1 - \delta)V_{\delta}(u))$ converge in $C(S)\times\mathcal{R}$ to (say) $(V, \beta)$. From (\ref{DP-disc}), we have
\begin{displaymath}
\bar{V}_{\delta}(x) + (1 - \delta)V_{\delta}(u) = \max\left(C\lambda +\delta\bar{V}_{\delta}(\alpha x + u), \ x\right).
\end{displaymath}
Passing to the limit along an appropriate subsequence as $\delta\uparrow 1$, we have
\begin{eqnarray}
V(x) + \beta &=& \max\left(C\lambda + V(\alpha x + u), \ x\right) \label{DP-av} \\
&=& \max_{v \in \{0, 1\}}\Big(vx + (1 - v)(\lambda + V(\alpha x + u))\Big). \nonumber \\
\label{DP-av2}
\end{eqnarray}
Then (\ref{DP-av}) is the desired dynamic programming equation for average reward.
We study important structural properties of the value function $V$ in the next section.\\

\section{Properties of the value function}

We begin with the following result.\\

\noindent \textbf{Lemma 2} The following statements hold: \\

\noindent $(1)$ $V$ is monotone increasing and convex with $V(u) = 0$;\\

\noindent $(2)$ The maximizer on the right hand side of (\ref{DP-av2}) is the optimal control choice at state $x$ and $\beta$ is the optimal reward.\\

\noindent \textbf{Proof:} Since monotonicity and convexity are preserved in pointwise limits, the first claim is immediate. For the second, let $v^*(x)$ denote the maximizer on the r.h.s.\ of (\ref{DP-av2}), any tie being settled arbitrarily. Then
under $\{v(t) = v^*(X(t)), \ t \geq 0\}$,
\begin{equation}
V(X(t)) + \beta  = k(X(t), v(t)) + V(X(t+1)). \label{temp}
\end{equation}
Summing (\ref{temp}) over $t = 1,2,\cdots,T$, and  dividing by $T$ on both sides, then letting $T\uparrow\infty$, we see that $\beta =$ the average reward under this control policy. On the other hand, for any other control sequence, the equality in (\ref{temp}) will be replaced by $\geq$, leading to the conclusion that $\beta \geq$
the corresponding average reward by an argument similar to the above. This imples the second claim. \hfill $\Box$

\ \\

Now define
\begin{eqnarray*}
B &:=& \{x \in S: C\lambda + V(\alpha x + u) > x\}, \\
B^c &:=& \{x \in S:  C\lambda + V(\alpha x + u) \leq  x\}.
\end{eqnarray*}
These are respectively the sets of passive and active states under subsidy $\lambda$.\\

Recall the stopping time $\tau :=$ the time of first crawl. Suppose $\tau < \infty$. (The case $\tau = \infty$ corresponds to `\textit{never crawl}' which we consider separately below.) Under optimal policy, iterating equation (\ref{DP-av}) $\tau$ times leads to
\begin{displaymath}
V(x) = (C\lambda - \beta)\tau + \left[\alpha^{\tau}x + \left(\frac{1 - \alpha^{\tau}}{ 1 - \alpha}\right)u - \beta\right].
\end{displaymath}
Under any other policy, we would likewise obtain
\begin{displaymath}
V(x)\geq (C\lambda - \beta)\tau + \left[\alpha^{\tau}x + \left(\frac{1 - \alpha^{\tau}}{ 1 - \alpha}\right)u - \beta\right].
\end{displaymath}
Thus we have the explicit representation for $V$ given by
\begin{displaymath}
V(x) =\max\left[(C\lambda - \beta)\tau + \left[\alpha^{\tau}x + \left(\frac{1 - \alpha^{\tau}}{ 1 - \alpha}\right)u - \beta\right] \right] ,
\end{displaymath}
where the maximum is over all admissible control sequences. In particular, this implies:\\

\noindent \textbf{Lemma 3} Equation (\ref{DP-av}) has a unique solution.\\

Finally, we have the key lemma:\\

\noindent \textbf{Lemma 4} The above problem is Whittle indexable.\\

\noindent \textbf{Proof:} \ Since $V$ is monotone increasing and convex, the map
\begin{displaymath}
x \mapsto x - V(\alpha x + u)
\end{displaymath}
is concave and hence the set $B$ increases monotonically from $\phi$ to $S$ as $\lambda$
increases from $-\infty$ to $\infty$. The claim now follows from the definition of Whittle indexability. \hfill $\Box$

\ \\

We shall now eliminate some irrelevant situations.\\

\begin{enumerate}
\item If $u^* \in B$, i.e., the optimal action at $u^*$ is $0$, then $u^*$ is a fixed point of the optimally controlled dynamics and the corresponding cost is $C \lambda$. Then $\beta = C\lambda$ and it is optimal to be passive at all states, i.e., $B = [u, u^*]$, $B^c = \phi$, and
\begin{equation}
\lambda \geq \lambda_m := \max_{x \in [u, u^*]}(x - V(\alpha x + u))/C. \label{high}
\end{equation}

\item If $u \in B^c$, then from (\ref{DP-av}), $0 + \beta = u + 0$, i.e., $\beta = u$ and it is optimal to crawl when at $u$. Then $u$ is a fixed point of the controlled dynamics and it is optimal to be active at all states, i.e., $B^c = [u, u^*]$, $B = \phi$, and
\begin{equation}
\lambda \leq \lambda_M := \min_{x \in [u, u^*]}(x - V(\alpha x + u))/C. \label{low}
\end{equation}
\end{enumerate}
Note that since constant policies $v(t) \equiv 0$ and $v(t) \equiv 1$ lead to costs $C\lambda$ and $u$ resp.,  $\beta \geq (C\lambda)\vee u$ always and $\beta > (C\lambda)\vee u$ for $\lambda \in (\lambda_m, \lambda_M)$. For each $\lambda$ in $(\lambda_m, \lambda_M)$, both $B, B^c$ are non-empty and there exists an $a \in (u, u^*)$ for which the choice of being active or passive is equally desirable. Furthermore, this $a$ is an increasing function of $\lambda$  by Lemma 4. Inverting this function, we have $\gamma(x) :=$ the value of $\lambda$ at which the active and passive become equally desirable choices, as an increasing function of $x \in (u, u^*)$. \\

\noindent \textbf{Lemma 5} The sets $B, B^c$ are of the form $[u, a), [a, u^*]$ for some $a \in [u, u^*]$.\\

\noindent \textbf{Proof:}  Since $V$ is convex, one of the following two must hold:
\begin{enumerate}
\item For some $a_2 > a_1$, $B = [u, a_1)\cap(a_2, u^*]$ and $B^c = [a_1, a_2]$, or,

\item for some $a$, $B = [u, a), B^c = [a, u^*]$.
\end{enumerate}
However, since at $u^*$ the optimal action is to crawl, we conclude that $u^* \in B^c$
and only the second possibility can occur. \hfill $\Box$

\ \\

\noindent \textbf{Corollary 1} The map $x \mapsto x - V(\alpha x + u)$ is monotone non-decreasing on $[u, u^*]$. \\

\section{Derivation of Whittle index}

Consider the situation when $\lambda = \gamma(x)$ for a prescribed $x \in (u, u^*)$. It is clear that after the first crawl when the process is reset to $u$, the optimal $X(t)$ becomes periodic: not crawling and increasing till it hits
 $B^c$ and
then crawling - thereby  being reset to $u$ - to repeat the process. Since finite initial patches do not affect the long run average reward, we may then take $X(0) = u$. Define $\eta(x) = \min\{t : X(t) \in B^c\}$. Then
\begin{eqnarray}
X(\eta(x)) &=& (1 - \alpha^{\eta(x)})u^* \label{hit} \\
\Longrightarrow \eta(x) &=& \left\lceil\log_\alpha^+\left(1 - \frac{x}{u^*}\right)\right\rceil, \label{eta}
\end{eqnarray}
where $\log_{\alpha}^+ x = \log_{\alpha}xI\{x > 0\}$. Since the long run average cost is equal to
the average over one period, we can write
\begin{equation}
\beta = \frac{C\lambda(\eta(x) - 1) + X(\eta(x))}{\eta(x)}, \label{beta}
\end{equation}
where $\eta(x)$ is given by (\ref{eta}) and $X(\eta(x))$ is given by (\ref{hit}). \\

We now revert to using  the index $i$ to identify the source being referred to. In particular, $\beta_i, \lambda_i$ will refer to the optimal reward, resp.\ Lagrange multiplier, for the $i$th decoupled problem. Our main result is:\\

\noindent \textbf{Theorem 1} The Whittle index for our problem is given by
\begin{displaymath}
\gamma_i(x) := \frac{1}{C_i}\left[\eta_i(x)((1 - \alpha_i)x - u_i) + \left(\frac{1 - \alpha_i^{\eta_i(x)}}{1 - \alpha_i}\right)u_i\right],
\end{displaymath}
where
\begin{displaymath}
\eta_i(x) := \left\lceil\log^+_{\alpha_i}\left(\frac{u_i - (1 - \alpha_i)x}{u_i}\right)\right\rceil.
\end{displaymath}
Therefore the index policy is to crawl at time $t \ (= mT$ for some $m \geq 0)$ the top $M$ sources according to decreasing values of $\gamma_i(X_i(t))$, or alternatively, choose a number of top sources for
the constraint to be reached. \\

\noindent \textbf{Remark:} Note that if an  arm (say, $i$th) is crawled even once, the corresponding state process $\{X_i(t)\}$ takes only discrete values thereafter. These depend on $\alpha_i$ and $u_i$ alone. In fact this is also true for an arm that is never crawled, except that the discrete values taken will also depend on the initial condition. Therefore we need restrict attention to only these values of $x$ for the argument of $\gamma_i(\cdot)$. This results in a further simplification of the index formula, to
$$
\gamma_i(x) = \frac{1}{C_i}\left(\eta_i((1 - \alpha_i)x - u_i) + x\right),
$$
where $\eta_i(x)$ is as before, but the argument $x$ of both $\gamma_i$ and $\eta_i$ is now restricted to the aforementioned discrete set.\\

\noindent \textbf{Proof:} \ We drop the subscript $i$ for notational convenience.
For $x \in B^c$, (\ref{DP-av}) leads to $V(x) = x - \beta$. Also, for $x' := \alpha x + u$,
\begin{eqnarray*}
x \leq u^* &=& \frac{u}{1 - \alpha}\\
 \Longrightarrow x' &=& \alpha x + u \\
 &\geq& \alpha x + (1 - \alpha)x \\
 \Longrightarrow x' &\geq& x \\
 \Longrightarrow x' &\in& B^c \ \ (\mbox{by Lemma 5}) \\
 \Longrightarrow V(x') &=& x' - \beta.
\end{eqnarray*}
Combining this with (\ref{DP-av}) and the definition of Whittle index implies that for our problem it is
\begin{equation}
\gamma_i(x) = \frac{(1 - \alpha_i)x - u_i + \tilde{\beta}_i(x)}{C_i}, \label{index}
\end{equation}
where by virtue of (\ref{beta}), $\tilde{\beta}_i(x) :=$ the optimal cost if one were to set $\lambda_i = \gamma_i(x)$. The latter is given by:
\begin{displaymath}
\tilde{\beta}_i(x) :=  \frac{1}{\eta_i(x)}\Big\{C_i\gamma_i(x)(\eta_i(x) - 1) + \left(1 - \alpha_i^{\eta_i(x)}\right)u^*_i\Big\}.
\end{displaymath}
where
\begin{displaymath}
\eta_i(x) := \left\lceil\log^+_{\alpha_i}\left(\frac{u_i - (1 - \alpha_i)x}{u_i}\right)\right\rceil.
\end{displaymath}
Substituting this back into (\ref{index}), one gets a linear equation for $\gamma_i(x)$ that can be solved to evaluate $\gamma_i(x)$ as
\begin{displaymath}
\gamma_i(x) := \frac{1}{C_i}\left[\eta_i(x)((1 - \alpha_i)x - u_i) + \left(\frac{1 - \alpha_i^{\eta_i(x)}}{1 - \alpha_i}\right)u_i\right].
\end{displaymath}
This completes the proof.
\hfill $\Box$

\medskip

\section{Stochastic case}

We now consider the fully stochastic situation when traffic at each source is observed as a random variable.
In fact one could also consider mixed situations when some sources are observed and others are not.
As we shall see, the development closely mimics the foregoing and the Whittle index is actually the same.

The stochastic system dynamics can be described as follows:
Let $\{\tau^i_n\}$ denote the successive arrival times of content at source $i$, with utilities $\{\xi^i_n\}$, resp.
The net utility added to source $i$ during $k$-th epoch will be
\begin{displaymath}
U_i(k) := \sum_{\tau^i_n \ : \ (k-1)T \leq \tau^i_n < kT}\xi^i_ne^{-\mu_i(kT - \tau^i_n)}.
\end{displaymath}
The system state at time $(k+1)T$ is then
\begin{eqnarray}
X_i(k+1) &=& \alpha_iX_i(k) + U_i(k+1) \ \ \ \mbox{if no crawl}, \nonumber \\
&=& U_i(k+1) \ \ \ \ \ \ \ \ \ \ \ \ \ \ \ \ \mbox{if crawled}. \label{eq:stochdyn}
\end{eqnarray}
We define the average reward as
\begin{displaymath}
\limsup_{t\uparrow\infty}\sum_{i=1}^N\frac{1}{t}\sum_{m=0}^tE[r(X_i(t), v_i(t))],
\end{displaymath}
which we seek to maximize subject to the constraint
\begin{displaymath}
\limsup_{t\uparrow\infty}\frac{1}{t}\sum_{i=1}^NC_iE[v_i(t)] = M.
\end{displaymath}
The discounted value function
\begin{displaymath}
V_{\delta}(x) := \sup_{\{v(t)\}, X(0) = x}E\left[\sum_{t=0}^{\infty}\delta^tk(X(t), v(t))\right]
\end{displaymath}
then satisfies the dynamic programming equation
\begin{equation}
V_{\delta}(x) = \max\left(C\lambda + \delta\int V_{\delta}(\alpha x +  u)\varphi_i(du), \ x + \delta \int V_{\delta}(u)\varphi_i(du)\right), \label{stochDP1}
\end{equation}
where $\varphi_i$ is the law of $U_i(t) \ \forall t$.\\

\noindent \textbf{Lemma 5} The conclusions of Lemma 1 continue to hold.\\

\noindent \textbf{Proof:} The first claim follows as before from the cited results of \cite{Lasserre}. For the second, let $X(t), X'(t)$ be as in the proof of Lemma 1 (2). Then
\begin{eqnarray*}
V_{\delta}(x') - V_{\delta}(x) &\leq& E\left[\sum_{t=0}^{\infty}\delta^m\left(k(X'(t), v(t)) - k(X(t), v(t))\right)\right] \\
&\leq& E\left[(\alpha\delta)^{\tau}\right](x' - x).
\end{eqnarray*}
The Lipschitz property follows as before. Next let $X(t), X'(t)$ be as in the proof of Lemma 1 (3). Taking expectations in (\ref{av-ineq}) followed by a supremum over all admissible controls proves monotonicity. 
Convexity follows as in the deterministic case. \hfill $\Box$

\ \\

The `vanishing discount' argument of Section 4 can now be used to establish the average cost dynamic programming equation
\begin{equation}
V(x) + \beta = \max(C\lambda +\int V(\alpha x + u)\varphi(du), \ x). \label{stochDPav}
\end{equation}
Monotonicity and convexity of $V$ follows as in Lemma 2. Equation (\ref{temp}) gets modified to
\begin{displaymath}
E[V(X(t))] + \beta  = E[k(X(t), v(t))] + E[V(X(t+1))],
\end{displaymath}
from which the optimality of
\begin{displaymath}
v^*(x) \in \mbox{Argmax}_v\left(vx + (1 - v)(\lambda + \int V(\alpha x + u)\varphi(du))\right),  \quad x \in S,
\end{displaymath}
follows by arguments analogous to those of Lemma 2. Furthermore, $V$ can be shown to be the unique solution of (\ref{stochDPav}) by establishing the explicit representation
\begin{displaymath}
V(x) = \max E\left[(C\lambda - \beta)\tau + \alpha^{\tau}x + \sum_{t=0}^{\tau} \alpha^{\tau - t}U(t) - \beta\right],
\end{displaymath}
 where the maximum is over all admissible control sequences. Thus, Whittle indexability follows as before. Define
 \begin{displaymath}
 \Xi(x) := E\left[\sum_{t=0}^{\eta(x)} \alpha^{\eta(x) - t}U(t) \Big| X(0) = x\right].
 \end{displaymath}
 The definitions of $B, B^c$ change to
 \begin{eqnarray*}
B &:=& \{x \in S: C\lambda + \int V(\alpha x + u)\varphi(du) > x\}, \\
B^c &:=& \{x \in S:  C\lambda + \int V(\alpha x + u)\varphi(du) \leq  x\}.
\end{eqnarray*}
Let $\eta_m, m \geq 1,$ denote the successive visits to $B^c$, i.e., the crawling times. Then
\begin{displaymath}
X(\eta_{m+1}) = \sum_{t=\eta_m}^{\eta_{m+1} - 1}\alpha^{\eta_{m+1} - t}U_t, \quad m \geq 1.
\end{displaymath}
As before, we may assume that $\eta_1(x) = 0$. Then the expression (\ref{eta}) for $\eta_2(x)$ will continue to hold. We denote it by $\eta(x)$ as before for notational convenience.  Therefore
\begin{eqnarray*}
\beta(x) &=& \frac{C\lambda(\eta(x) - 1) + E[X(\eta(x))]}{\eta(x)} \\
&=& \frac{C\lambda(\eta(x) - 1) + (1 - \alpha^{\eta(x)})u^*}{\eta(x)}
\end{eqnarray*}
as before. Hence the conclusions of Theorem 1 continue to hold.

\section{Numerical examples}

\begin{figure}[ht]
      \centering
      \vspace{-4cm}
      \includegraphics[scale=0.4]{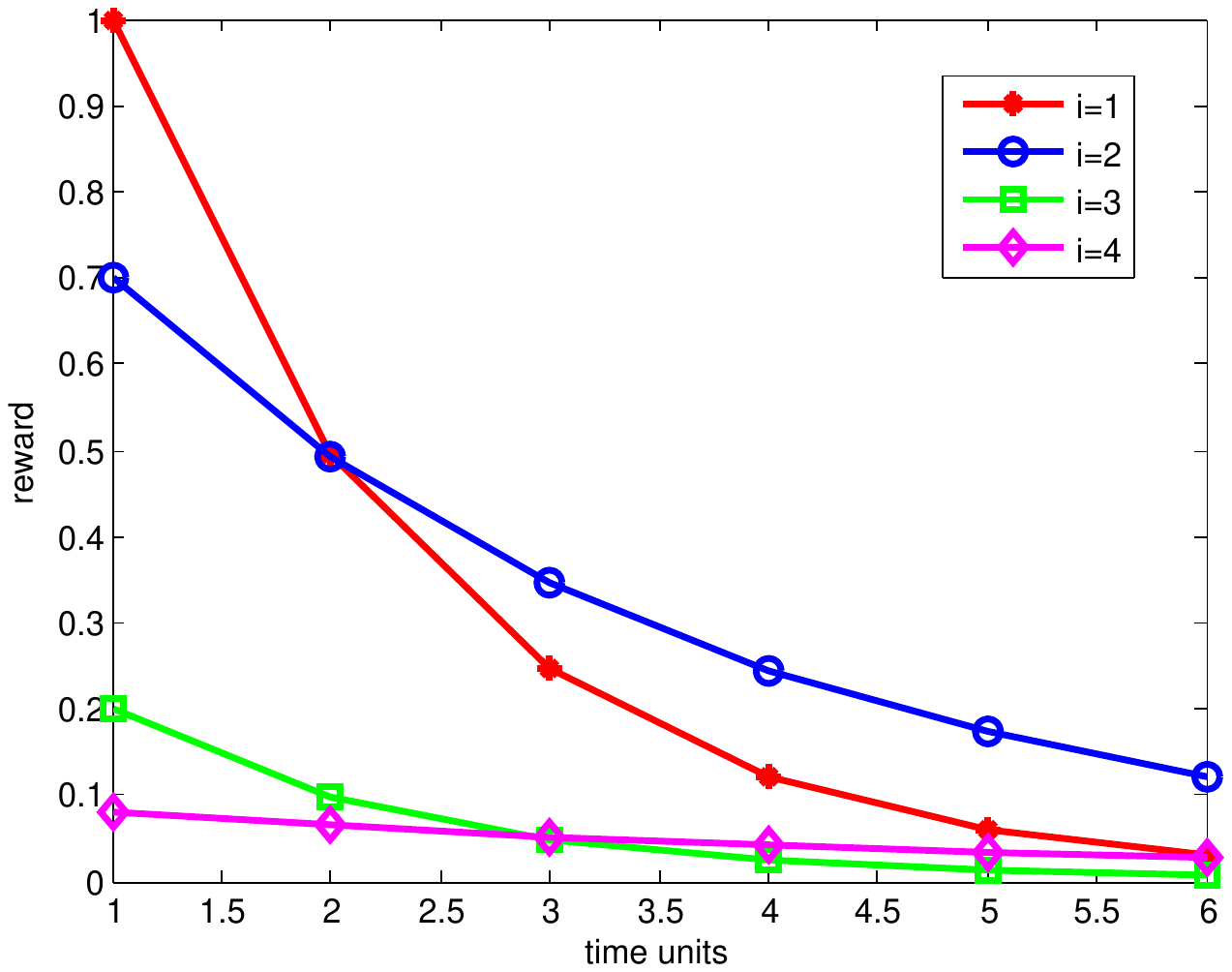}
      \vspace{-4cm}
      \caption{Content value as a function of time.}
      \label{figrewardstime}
   \end{figure}

Let us illustrate the obtained theoretical results by numerical examples.
There are four information sources with parameters given in Table~\ref{tabexample}.
Without loss of generality, we take the crawling period $T=1$.
One can see how the user interest decreases over time for each source in Figure~\ref{figrewardstime}.
The initial interest in the content of sources 1 and 2 is high, whereas the initial
interest in the content of sources 3 and 4 is relatively small. The interest in
the content of sources 1 and 3 decreases faster than the interest in the content
of sources 2 and 4.

\begin{figure}[ht]
      \centering
      \vspace{-4cm}
      \includegraphics[scale=0.4]{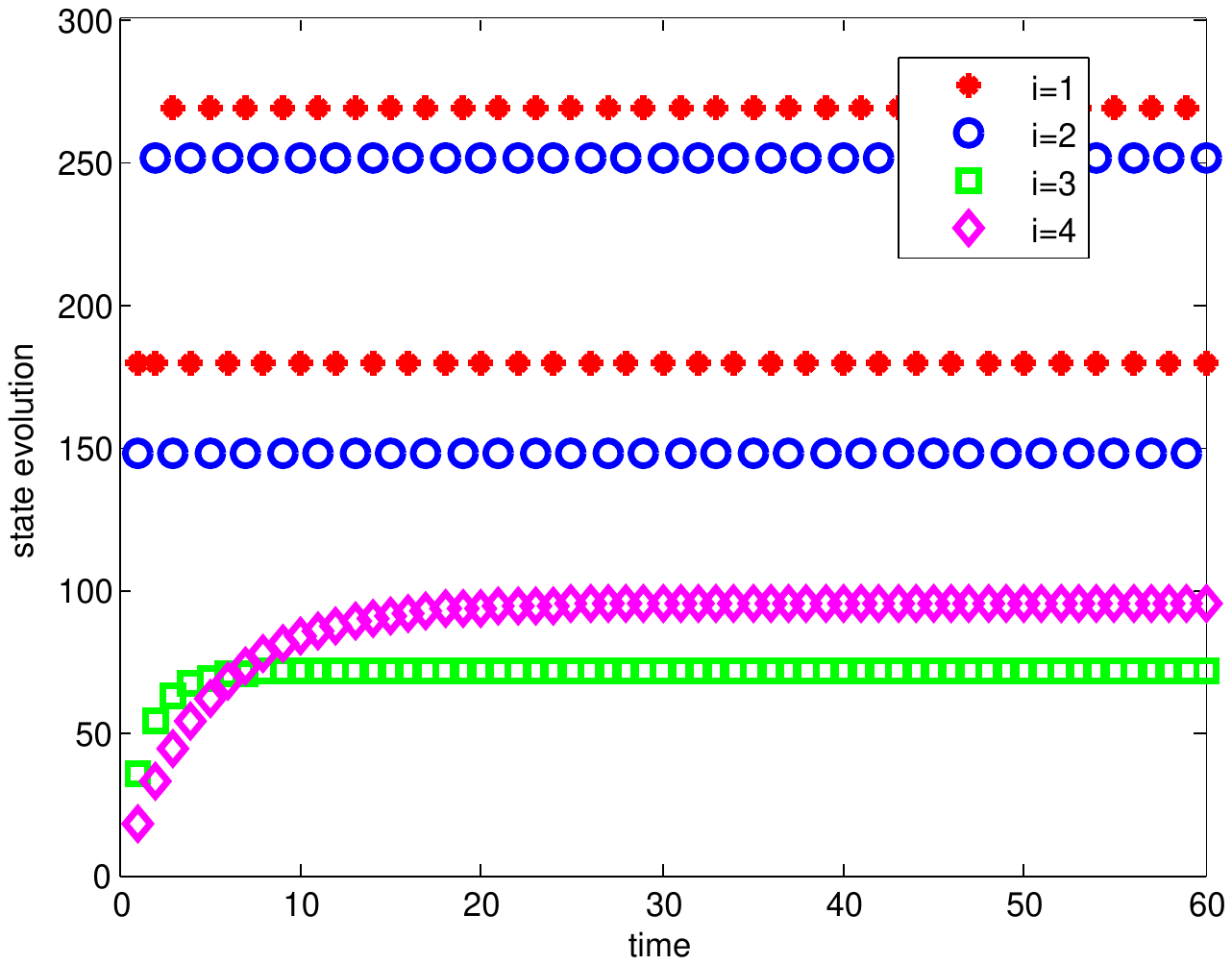}
      \vspace{-4cm}
      \caption{The case of $M=1$.}
      \label{figM1}
   \end{figure}

In Figure~\ref{figM1} we show the state evolution of the bandits (information sources)
under the constraint that on average the crawler can visit only one site per crawling
period $T$, i.e., $M=1$. The application of Whittle index results in periodic crawling
of sources 1 and 2, crawling each with period two. Sources 3 and 4 should be never crawled.
Note that if one greedily crawls only source 1, he obtains the average reward $179.79$.
In contrast, the index policy involving two sources results in the average
reward $254.66$.

\begin{figure}[ht]
      \centering
      \vspace{-4cm}
      \includegraphics[scale=0.4]{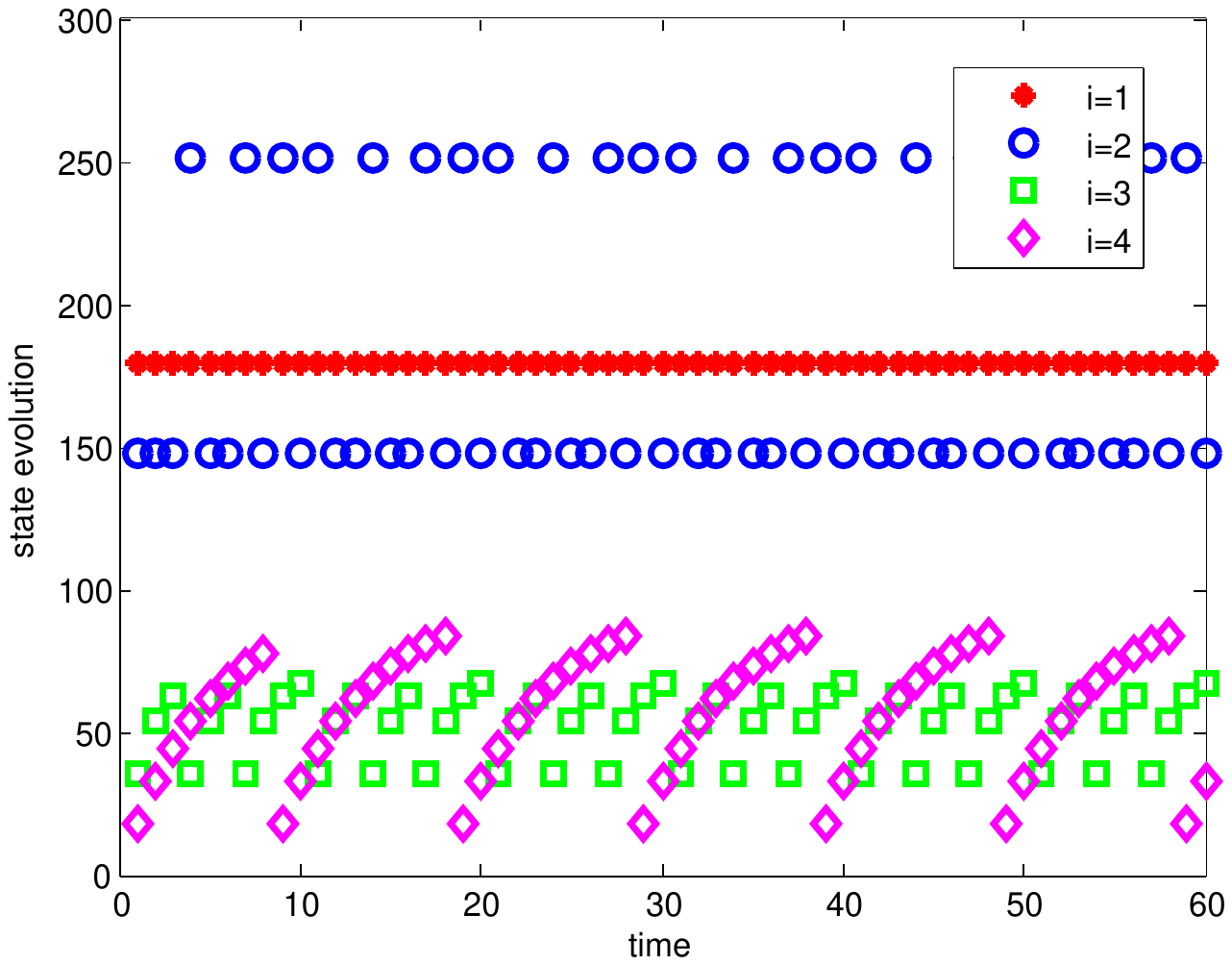}
      \vspace{-4cm}
      \caption{The case of $M=2$.}
      \label{figM2}
   \end{figure}

\begin{figure}[ht]
      \centering
      \vspace{-4cm}
      \includegraphics[scale=0.4]{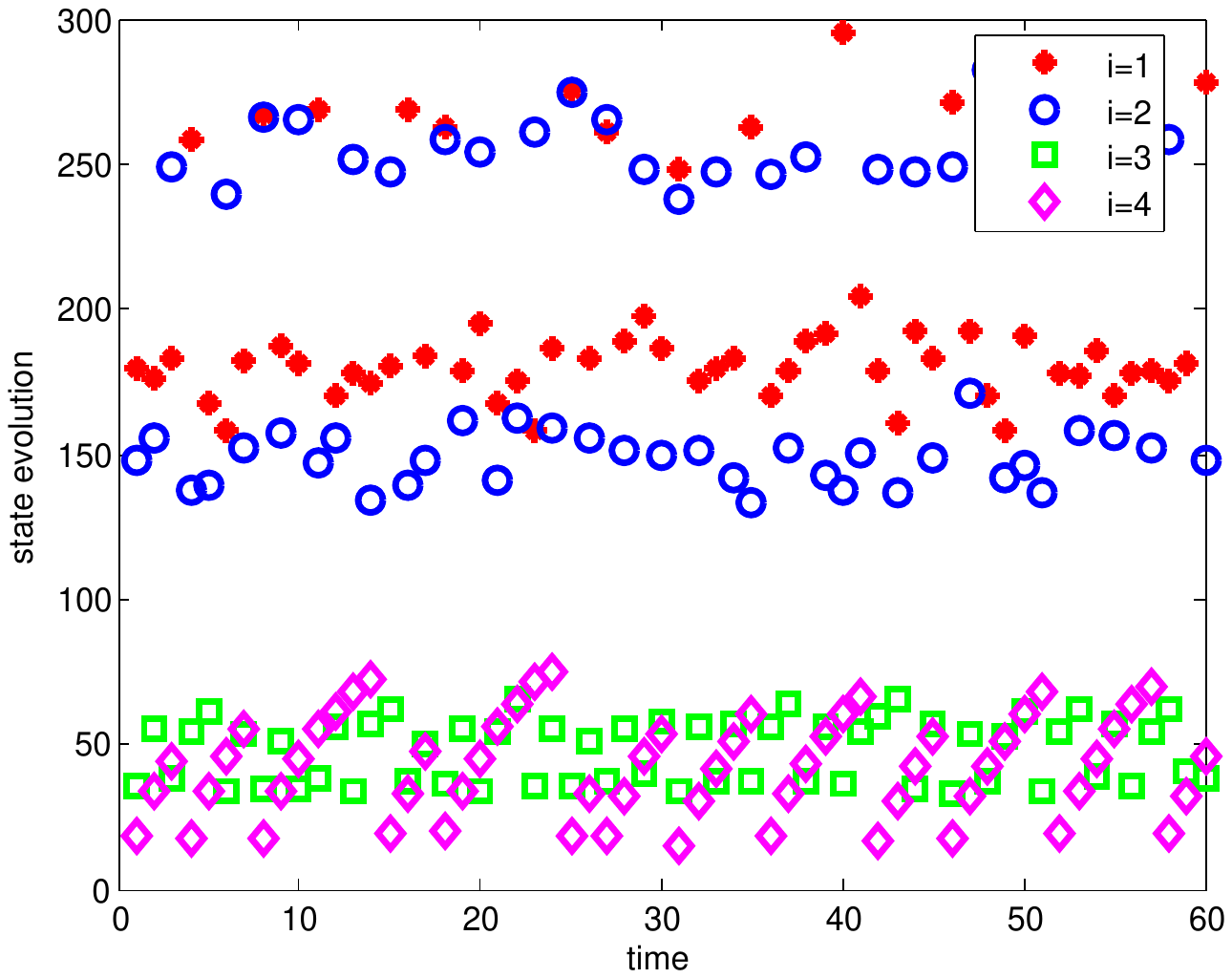}
      \vspace{-4cm}
      \caption{The case of $M=2$ (stochastic model).}
      \label{figM2s}
   \end{figure}

In Figure~\ref{figM2} we show the state evolution of the bandits under the constraint
that on average the crawler can visit two information sources per crawling period, i.e., $M=2$.
It is interesting that now the policy becomes much less regular. Source 1 is always crawled.
Sources 2 and 3 are crawled in a non-trivial periodic way and sources 4 is crawled periodically
with a rather long period. Now in Figure~\ref{figM2s} we present the state evolution of the
stochastic model with dynamics (\ref{eq:stochdyn}). As one can see, in the stochastic setting source 1
is crawled from time to time.

\begin{table}[ht]
\caption{Data for numerical example}
\label{tabexample}
\begin{center}
\begin{tabular}{|c|c|c|c|c|}
\hline
$i$ & 1 & 2 & 3 & 4\\
\hline
$\bar{\xi}_i$ & 1.0 & 0.7 & 0.2 & 0.08 \\
\hline
$\mu_i$ & 0.7 & 0.35 & 0.7 & 0.21 \\
\hline
$\Lambda_i$ & 250 & 250 & 250 & 250 \\
\hline
\end{tabular}
\end{center}
\end{table}


\section{Conclusions and future works}

We have formulated the problem of crawling web sites with ephemeral content as
an average reward optimal control problem and have shown that it is indexable.
We have found that the Whittle index has a very simple form, which is important
for efficient practical implementations. The numerical example demonstrates that
the Whittle index policies, unlike the policies suggested in \cite{LOSS11}, do not
generally have a trivial periodic structure. The proposed approach can also be used
in the cases when some states are observed. In such cases, the Whittle
index will act as a self-tuning mechanism. We are currently working on the adaptive
version when some parameters (e.g., the rate of new information arrival) need to be
estimated online. One more interesting future research direction is to add to the
model the dynamics of the indexing engine.

\section*{Acknowledgments}

The authors gratefully acknowledge the discussions with Liudmila O. Prokhorenkova and
Egor Samosvat from Yandex during the preparation of the manuscript.

\newpage
\tableofcontents

\end{document}